\documentclass[11pt,twocolumn,prb,superscriptaddress,reprint]{revtex4-1}

\usepackage{graphicx,amsmath,amssymb,color}
\usepackage{tabularx, ctable}

\usepackage{ulem}
\usepackage{bm}
\usepackage{siunitx}
\usepackage{gensymb}
\usepackage{xr}
\usepackage[breaklinks=true,colorlinks,allcolors=blue]{hyperref}

\newcommand{\be}{\begin{eqnarray}}
\newcommand{\ee}{\end{eqnarray}}

\begin{document}

\title{{Graphene Zero-Bias Sub-Terahertz Turnkey Detector with Above 43 GHz Bandwidth}}

\author{E.I. Titova$^{*}$}
\affiliation{Programmable Functional Materials Lab,
Center for Neurophysics and Neuromorphic Technologies, Moscow, 121205, Russia}
% \affiliation{Center for Photonics and 2D Materials, Moscow Institute of Physics and Technology, Dolgoprudny, 141700, Russian Federation}
\affiliation{Center for Advanced Studies, Kulakova str, Moscow}

\author{A. Titchenko}
\affiliation{National Research University Higher School of Economics, Moscow, 101000}
\affiliation{Moscow Pedagogical State University, Moscow 119991}

\author{M.~Titova}
\affiliation{Programmable Functional Materials Lab,
Center for Neurophysics and Neuromorphic Technologies, Moscow, 121205, Russia}
\affiliation{Center for Advanced Studies, Kulakova str, Moscow}

\author{K. Shein}
\affiliation{National Research University Higher School of Economics, Moscow, 101000}
\affiliation{Moscow Pedagogical State University, Moscow 119991}

\author{A. Kuksov}
\affiliation{Programmable Functional Materials Lab,
Center for Neurophysics and Neuromorphic Technologies, Moscow, 121205, Russia}

\author{A. Sobolev}
% \affiliation{Moscow Institute of Physics and Technology (National Research University)}
\affiliation{Center for Advanced Studies, Kulakova str, Moscow}

\author{S. Zhukov}
% \affiliation{Moscow Institute of Physics and Technology (National Research University)}
\affiliation{Center for Advanced Studies, Kulakova str, Moscow}

\author{K.~Kapralov}
\affiliation{Center for Advanced Studies, Kulakova str, Moscow}

\author{E. Zhukova}
% \affiliation{Moscow Institute of Physics and Technology (National Research University)}
\affiliation{Center for Advanced Studies, Kulakova str, Moscow}

\author{Ya. Gershtein}
\affiliation{Programmable Functional Materials Lab,
Center for Neurophysics and Neuromorphic Technologies, Moscow, 121205, Russia}
\affiliation{Center for Advanced Studies, Kulakova str, Moscow}

\author{M.~Kashchenko}
\affiliation{Programmable Functional Materials Lab,
Center for Neurophysics and Neuromorphic Technologies, Moscow, 121205, Russia}
% \affiliation{Center for Photonics and 2D Materials, Moscow Institute of Physics and Technology, Dolgoprudny, 141700, Russian Federation}
\affiliation{Center for Advanced Studies, Kulakova str, Moscow}

\author{A.~Shabanov}
\affiliation{Center for Advanced Studies, Kulakova str, Moscow}

\author{M. Kravtsov}
\affiliation{Department of Materials Science and Engineering, National University of Singapore, 117575, Singapore}

\author{L. Elesin}
\affiliation{Department of Materials Science and Engineering, National University of Singapore, 117575, Singapore}

\author{K. S. Novoselov}
\affiliation{Institute for Functional Intelligent
Materials, National University of Singapore, Singapore, 117575, Singapore}

\author{G. Goltsman}
\affiliation{Moscow Pedagogical State University, Moscow 119991}

\author{D. A. Svintsov}
% \affiliation{Center for Photonics and 2D Materials, Moscow Institute of Physics and Technology, Dolgoprudny, 141700, Russian Federation}
\affiliation{Center for Advanced Studies, Kulakova str, Moscow}

\author{I. Gayduchenko}
\affiliation{National Research University Higher School of Economics, Moscow, 101000}
\affiliation{Moscow Pedagogical State University, Moscow 119991}

\author{D. A. Bandurin$^{*}$}
\affiliation{Department of Materials Science and Engineering, National University of Singapore, 117575, Singapore}
\affiliation{Institute for Functional Intelligent
Materials, National University of Singapore, Singapore, 117575, Singapore}

\begin{abstract}

\textbf{High-frequency terahertz (THz) detectors are vital for the next-generation high-speed wireless communication systems. Graphene, with its high carrier mobility, broadband absorption, and weak electron-phonon coupling, offers great promise for ultra-fast THz photothermoelectric devices. Although graphene-based detectors in the infrared range have shown bandwidths above 500 GHz, extending their operation to the THz range is difficult because long-wavelength radiation does not efficiently couple to the small graphene area. To overcome this issue, THz antennas are often employed; however, their utilization typically limits system performance to only a few gigahertz due to parasitic effects. In this work, we present an antenna-coupled sub-THz graphene detector with a bandwidth exceeding 43 GHz. We optimized the detector design to minimize losses, match the antenna impedance to the 1 kOhm graphene channel, and maintain zero-bias operation. Importantly, we introduce a compact, turnkey packaged solution. Our results provide a practical route toward high-speed and low-power graphene THz detectors suitable for real-world communication and imaging applications.}

\begin{center}
$^{*}$ Correspondence to: titova.elenet@gmail.com;
dab@nus.edu.sg

\end{center}

\end{abstract}

\maketitle

\begin{figure*}[ht!]
  \centering\includegraphics[width=1\linewidth]{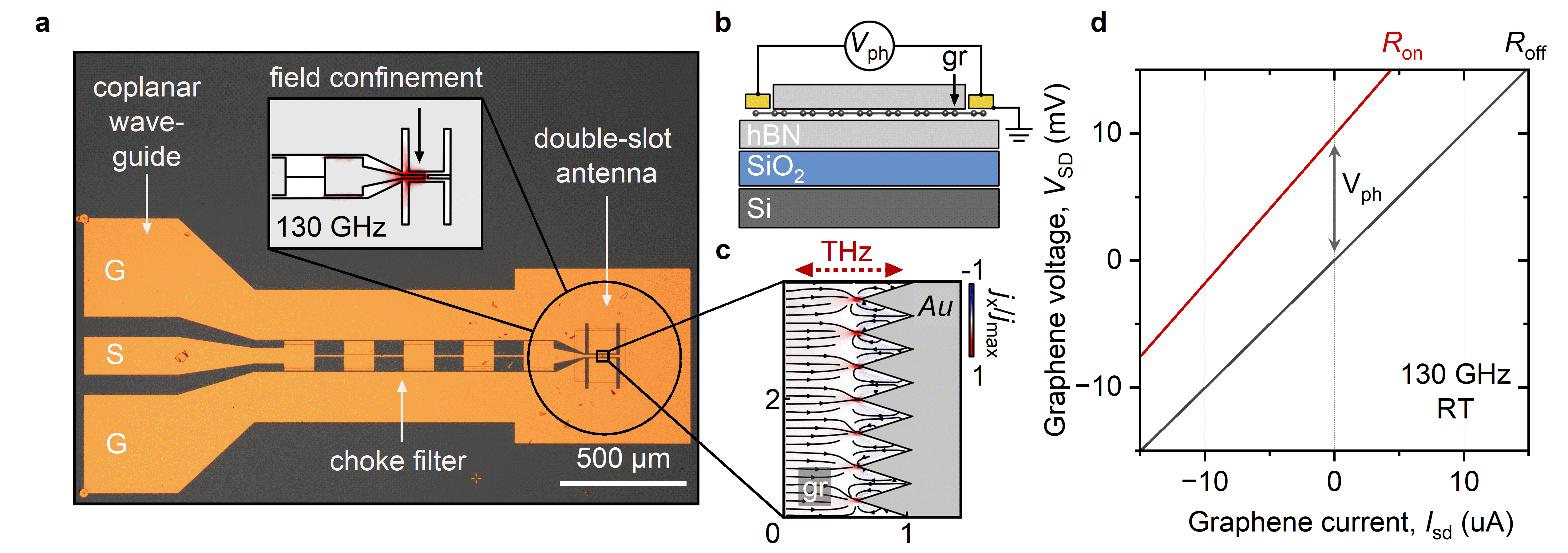}
    \caption{\textbf{Zero-bias graphene-based ultrafast THz detector.}
    % High‑impedance sub-THz antenna integrated with a coplanar waveguide for a fast graphene detector.}
    \textbf{(a)} Optical photograph of the graphene device consisting of a THz antenna, low‑pass choke filters, and a coplanar waveguide. 
    % The metal electrodes simultaneously serve as the source and drain contacts. 
    The photovoltage is readout between signal (S) and ground (G) electrodes. 
    % Scale bar: 500 $\mu$m. 
    Inset: Simulated electric field distribution in the antenna-coupled graphene detector under 130 GHz illumination, showing pronounced field enhancement in the graphene active region. 
    % Scale bar: 500 $\mu$m.
    % The right inset shows the asymmetric tooth structure adjacent to the graphene channel, designed to generate a zero‑bias photoresponse. Scale bar: 1 $\mu$m. 
    % The upper right inset shows schematic of the high-frequency graphene photodetector implemented with a silicon sphere lens.
    % The upper inset shows the dimensions of the double-slot antenna designed for 130 GHz.
    \textbf{(b)} Schematic cross-section of the hBN-encapsulated graphene detector, showing the graphene channel connected to signal and ground metal electrodes.
    \textbf{(c)} Simulation of zero-bias photocurrent generated by local field enhancement at the metal teeth, based on the photothermoelectric (PTE) model of photoresponse under 130 GHz illumination. The red arrow indicates the polarization direction of the incident THz radiation.The y-axis is scaled by a factor of two relative to the x-axis. Both axes are expressed in micrometers.
    % The inset shows the corresponding channel resistances $R = dV/dI$.
    \textbf{(d)}  Measured I–V curves of the graphene channel with (red) and without (black) 130 GHz THz illumination from a backward wave oscillator (BWO) in direct detection mode. A pronounced zero-bias photovoltage is evident.
    }
	\label{Fig1}
\end{figure*}

The rapidly increasing volume of information driven by Big Data, artificial intelligence, and virtual reality calls for the development of new high-performance electronic devices. This demand extends to wireless communications, where achieving higher data transfer rates and network capacity requires operation at progressively higher carrier frequencies. For example, peak data rates have grown from about 256 kbps in 2G networks operating near 1.8 GHz to approximately 50 Gbps in 5G systems using carrier frequencies of up to 50GHz~\cite{Jiang2024}. The ongoing transition toward 6G technology, which aims to deliver even greater capacity and speed, will therefore depend on detectors and receivers capable of efficient performance at frequencies of 100 GHz and beyond.

{Currently, the high-frequency signal detection in wireless technologies is achieved with active transistor-based amplifiers, the dominant technology represented by high electron mobility transistors~\cite{6G_HEMTs}. Among the drawbacks of this conventional approach are high energy-consumption and noise induced by bias currents. These can be mitigated with zero-bias detectors, among which III-V based Schottky diodes have achieved the best performance metrics~\cite{ZBD_technology}. Still, the Schottky diode technology also has limitations due to the large junction capacitance and large resistance, leading to the droop in efficiency  as frequencies approach the THz regime. Consequently, achieving fast and efficient response in the THz range remains a major challenge for traditional device architectures.}

Graphene, with its broadband absorption, high mobility, and ultrafast carrier dynamics, offers an attractive alternative for THz detectors~\cite{Vicarelli2012,DelgadoNotario2022,Muraviev2013,Castilla2019,Liu2021}. In most experimental arrangements, the photoresponse in graphene emerges via the hot-carrier photothermoelectric effect (PTE)~\cite{Tielrooij_PTE_grapehene_metal}. 
Owning to high energy of optical phonons in graphene, photoexcited hot electrons remain decoupled from the lattice~\cite{Low_cooling}, yielding a fast and pronounced PTE response. Following excitation, electrons thermalize on a femtosecond scale, with the hot Fermi distribution forming in $\approx$ 100 fs~\cite{Breusing2011}. Subsequent cooling of hot carriers via phonons then occurs over a few picoseconds~\cite{Pogna2021}, setting the intrinsic timescale for graphene-based photodetectors. Moreover, the all-in-plane design of these detectors yields low inter-electrode capacitance~\cite{Ultrafast_PD_capacitance}, enabling electrically measured extrinsic response times to approach the intrinsic detector limit.

% enabling extrinsic electrically measured response times to approach the intrinsic limit.
% Large magnitude of PTE is facilitated by high energy of graphene optical phonons and decoupling of hot carrier temperature from that of the lattice~\cite{Low_cooling}. 
% After excitation, electrons thermolized on a scale of femtoseconds: the time of hot Fermi distribution formation is of the order of 100 fs~\cite{Breusing2011}. Subsequent hot-carrier cooling on phonons occurs within approximately a few picoseconds~\cite{Pogna2021}. The latter define the intrinsic timescale of graphene-based photodetectors. Furthermore, the all-in-plane design of graphene detectors results in small inter-electrode capacitance~\cite{Ultrafast_PD_capacitance}, which should help achieve extrinsic electrically measured response times of the order of intristic response time. 
% Short extrinsic times in graphene detectors are further facilitated by small values of inter-electrode capacitance which, in turn, comes from all-in-plane detector arrangement~\cite{Ultrafast_PD_capacitance}.

% Carrier dynamics have been found to be ultrafast not only in high-quality exfoliated graphene but also in CVD-grown samples~\cite{Oum2014}, highlighting the strong potential of graphene-based detectors for commercial applications.

Experimental studies of fast graphene photodetectors have mostly focused on the near‑infrared (telecom) range. The highest reported bandwidth exceeds 500 GHz in optical‑to‑optical (O/O) measurements at 1550 nm~\cite{Koepfli2023}. Optical-to-electrical (O/E) measurements typically show lower frequency limits due to losses in interconnects and cables; however, high-frequency probe measurements at 1550 nm have demonstrated bandwidths of up to 75 GHz and connectorized packaged setups of up to 65 GHz~\cite{Wu2024}.

In principle, the almost frequency‑independent optoelectronic properties of graphene should enable the extension of such fast infrared devices to the THz range. However, in this regime the radiation wavelength (several millimeters) greatly exceeds the size of the graphene device (typically $\approx$ 10 $\mu$m), necessitating the use of special THz antennas to concentrate the radiation onto the active area. This introduces new challenges for high‑frequency characterization. Antennas placed near graphene without direct electrical contact (for example, used as a gate) significantly increase the capacitance and limit the detector speed — the bandwidth is typically restricted to a few GHz~\cite{Soundarapandian2026,Generalov2017}.  As demonstrated in Ref.~\cite{Soundarapandian2026}, removing the antenna can extend the bandwidth up to 40 GHz (measured using a high‑frequency probe station), although this improvement comes at the cost of a several‑orders‑of‑magnitude reduction in responsivity. Alternatively, using the antenna as direct electrical contacts for signal readout leads to impedance mismatch: while typical antenna impedance is below 50-100 Ohm, the graphene resistance near the charge‑neutrality point, where the photoresponse is strongest, is around 1 kOhm. This mismatch causes signal loss and limits the high‑frequency performance of graphene THz detectors. One possible solution is to reduce the graphene resistance using interdigitated finger contacts~\cite{Mueller2010}, but this approach is technologically complex and imposes constraints on the detector design. Here, we propose an alternative concept — employing a high‑impedance antenna matched to the graphene with a resistance of approximately 1 kOhm. The metal electrical contacts were designed as a multifunctional structure that integrates a THz antenna, choke filter, and on-chip coplanar waveguide. This design enables zero-bias photodetection via the asymmetric tooth profile patterned directly into the graphene active area.
% We specifically designed the metal contacts to the graphene area to serve as a THz antenna, choke filter, and on-chip coplanar waveguide.
% We demonstrate the packaged graphene detector integrated with Si lens lens design - tooth design, on-chip coplanar. 
% Zero‑bias photodetection was achieved through the asymmetric tooth design of the metal contacts to graphene. 

We demonstrate optoelectronic (O/E) measurements of a zero‑bias, room temperature sub-THz packaged graphene detector, showing a bandwidth exceeding 43 GHz, limited by the electrical spectrum analyzer. The frequency response was characterized in a heterodyne scheme by beating two sub‑THz sources and illuminating the graphene. To the best of our knowledge, this represents the highest demonstrated THz bandwidth for graphene-based photodetectors.

\begin{figure*}[ht!]
  \centering\includegraphics[width=1\linewidth]{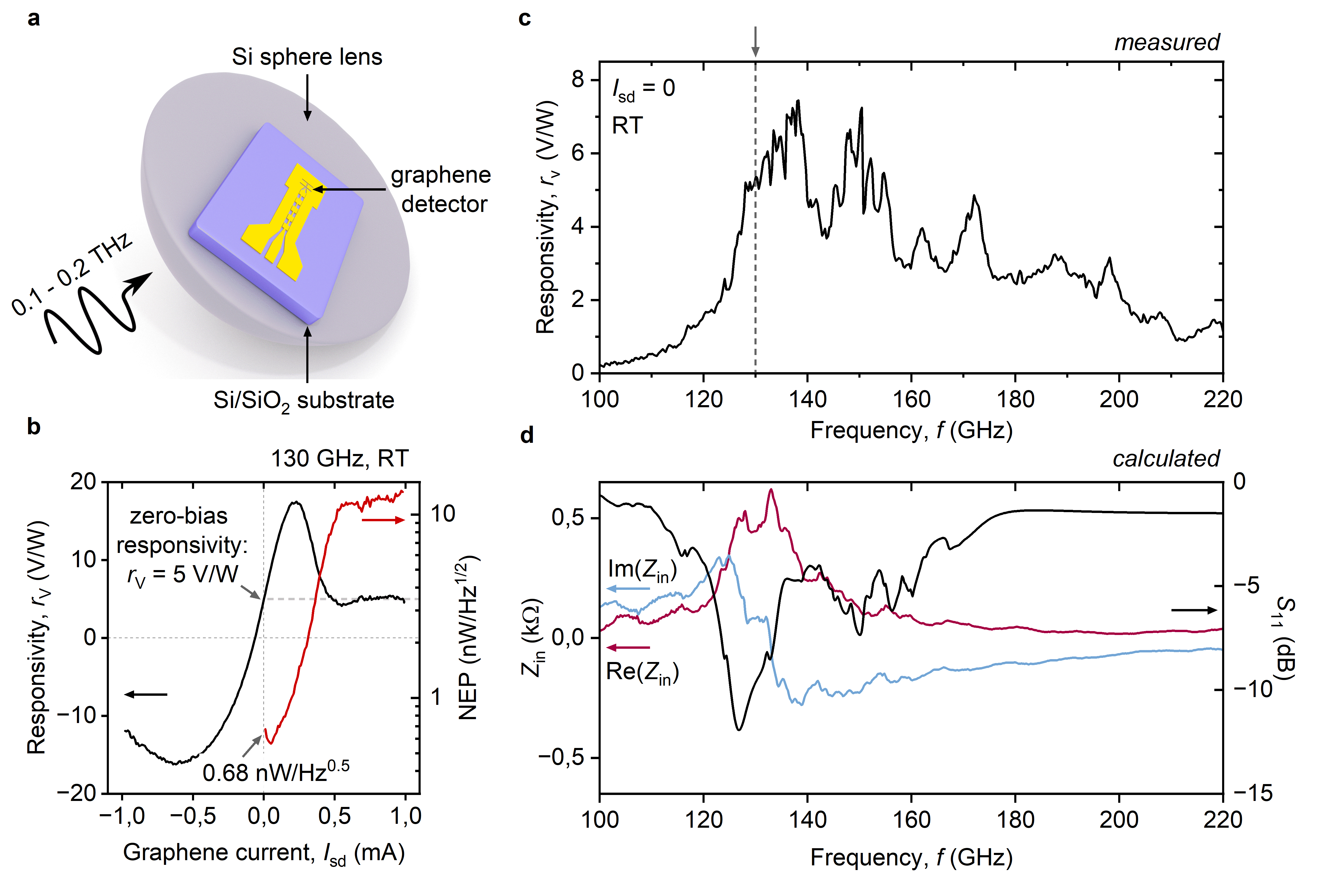}
    \caption{{\textbf{Direct detection characterization of the graphene detector.} 
    \textbf{(a)} Schematic of the detector mount: Silicon substrate with graphene detector positioned at the center of a silicon lens.
    \textbf{(b)} Detector responsivity ($r_V$) and noise-equivalent power (NEP) as a function of graphene current under 130 GHz illumination. $r_V = V_\mathrm{ph}/P_\mathrm{inc}$ and NEP $= s_V / r_V$ (with $s_V$ the voltage noise spectral density measured in 5 kHz bandwidth). At zero bias, $r_V = 5$ V/W and NEP $= 0.68$ nW/Hz$^{1/2}$.
    \textbf{(c)} Measured zero-bias photovoltage spectra of the graphene detector, recalculated to responsivity, over the 100–220 GHz frequency range. The dashed line indicates the 130 GHz frequency.
    \textbf{(d)} Calculated impedance ($Z_{in}$) and S-parameter of the sub-THz antenna loaded to 1 kOhm graphene. }}
	\label{Fig2}
\end{figure*}

\section*{Fast THz Graphene detector} 

The active element of our fast THz  detector was fabricated using exfoliated graphene encapsulated between two hBN flakes (see Fig.\ref{Fig1} a,b). The hBN–graphene–hBN stack was placed on a moderately p‑doped silicon substrate ($\approx$ 100 Ohm*cm) covered with a 300 nm thermal SiO$_2$ layer (see details in the Method section). To create the asymmetry required for zero‑bias photodetection, one of the graphene-metal electrodes was patterned into a tooth-like structure (see the grey contacts in Fig.\ref{Fig1} c and Fig.~S1), which has previously been shown to enable a zero-bias response in infrared photodetectors~\cite{Semkin2025}. The anticipated detection principle lies in local absorption enhancement at the structured metallic contact, leading to strong PTE at the adjacent metal-graphene junction. The PTE at the another junction, having the opposite sign, is minimized due to the absence of metal structuring and weaker radiation absorption.

% To achieve a zero-bias photodetection regime, we designed an asymmetric tooth-like electrode structure. 
% The sharp metallic edges serve as hotspots for local field enhancement. 
The photocurrent, calculated using the PTE rectification model (see calculation details in the Methods section), exhibits a unipolar flow from one electrode to the other (see Fig.~\ref{Fig1}c). As expected, the electric field is strongly enhanced near the sharp, tooth-like needle structures, resulting in a pronounced photocurrent response.
% Interestingly, unlike the case observed in the infrared range~\cite{Semkin2025}, this tooth-like configuration shows no reverse current, suggesting an improved rectification capability.

% {Zero-bias photoresponse is achieved through geometric asymmetry of the metal-graphene contacts. One electrode is patterned into a tooth-like structure, producing localized field enhancement at its sharp edges under illumination. This leads to asymmetric carrier heating and an imbalance of the photothermoelectric response at the two contacts, resulting in a finite photovoltage without external bias.}

The graphene THz detector was embedded in a central electrode of the coplanar line connecting two half-wavelength ($\lambda/2$)  slots that form a double-slot lens antenna~\cite{THz_antenna_Filipovic1993} (see Fig.\ref{Fig1} a).  Such an antenna is suitable for detectors with the typical impedance of about 50 to 100 Ohm~\cite{THz_antenna_Zhang2011}, but we designed our antenna for 1 k$\Omega$ impedance (see impedance spectra in Fig.\ref{Fig2} d) by setting the detector-to-slot distance to $\lambda/4$.
% as shown in Fig.~\ref{Fig1} a. 
This RF circuit enables DC bias application and intermediate-frequency signal extraction without any loss of the high-frequency power received by the antenna. 
% The input antenna impedance varies depending on the length of the coplanar line section. This impedance can exceed 1 kOhm when the distance between the detector and the slots is lambda quarter, 
% allow frequency circuit is designed for applying DC-bias to the detector and extracting the intermediate frequency signal without any loss of the high-frequency power received by the antenna. 
The circuit also contains a choke filter consisting of several alternating $\lambda/4$ segments of the coplanar line. The high-frequency electric field is concentrated in the narrow 6 $\mu$m gaps between the central and ground conductors of the low-impedance segments, which may turn the low frequency circuit into a shunting resistor if the conductivity of the underlying substrate is too high. 
% In order to control the back-gating voltage our graphene detector is fabricated on an oxidized silicon substrate with the DC electric conductivity of about 100 Ohm*cm, which also allows the choke filter work as an RF transmission line and introduces less than 10$\%$ loss into the antenna performance.

% The on-chip RF-circuit comprises a THz antenna, an integrated choke filter, and biasing pads  (see Fig.~\ref{Fig1}). THz antenna was calculated as a high‑impedance double-slot antenna for a load impedance of 1 kOhm to match the graphene channel resistance. To minimize noise in the structure, on‑chip low‑pass filters were implemented to suppress frequencies above 100 GHz, preventing them from propagating through the electrical contacts. The filtered metallization was then coupled into a coplanar waveguide.

\begin{figure*}[ht!]
  \centering\includegraphics[width=1\linewidth]{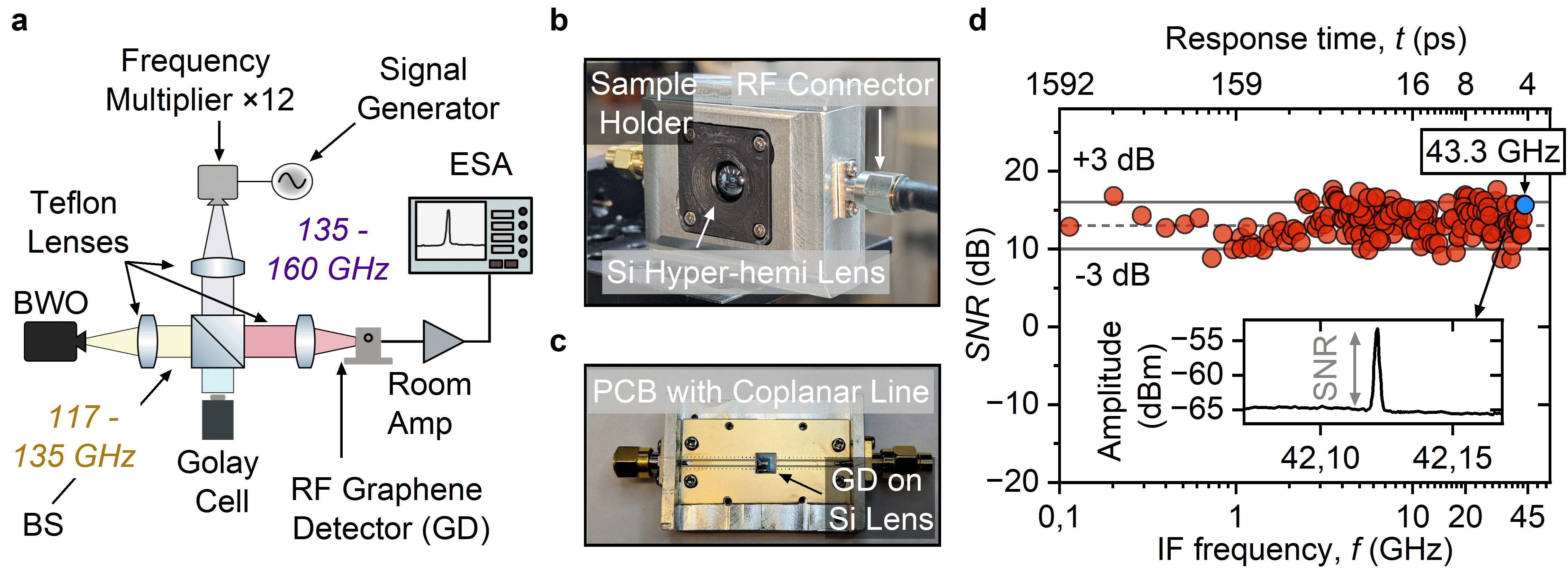}
    \caption{\textbf{Electrically measured response time and bandwidth (BW) of the THz graphene detector. }
    % Electrically measured (O/E) bandwidth in the THz graphene detector.}
    \textbf{(a)} Heterodyne measurement scheme. Two THz sources are combined using a beamsplitter (BS) to illuminate the detector, with the signal recorded by an electrical spectrum analyzer (ESA) at the intermediate frequency $\rm{IF} = |f_{\mathrm{source_1}} - f_{\mathrm{source_2}}|$. A Golay cell monitors THz power.
    \textbf{(b)} 
    % The photograph of the packaged graphene-based detector assembly with integrated electrical connectors.
    Photograph of the packaged detector module with integrated RF connector and Si hyper-hemispherical lens.
    \textbf{(c)} 
    % The photograph of high-frequency printed circus board (PCB) integrated with graphene detector on Si lens.
    High-frequency PCB assembly with coplanar waveguide contacting the graphene device.
    \textbf{(d)} Measured detector signal‑to‑noise photoresponse as a function of the intermediate frequency. The flat dependence indicates a bandwidth well above 43 GHz (corresponding to a response time of $t = 1/(2\pi f) \approx 3.7 ps$), limited by the measurement setup. Inset: Signal peak recorded by electrical spectrum analyzer at 42.12 GHz.
    }
	\label{Fig3}
\end{figure*}

{Connecting a high-frequency detector to measurement equipment is a non-trivial task. Interconnectors must be carefully designed, as signal losses and parasitic effects can severely distort or degrade the detector’s response.} Electrical connections of our detector to external circuitry were made via a specially designed high‑frequency printed circuit board (PCB) using wire bonding. A custom metal sample holder (see photo in Fig.~\ref{Fig3} b) was designed to simultaneously mount the PCB, sample, and silicon lens — with the sample positioned at the lens center — while also functioning as an integrated Faraday cage to shield the system from external electromagnetic interference.

\section*{Device Characterization} 

We first verified the device characteristics through DC electrical measurements. The measured DC IV curve shows a graphene resistance of about 1 kOhm, which increases by roughly 100 Ohm under THz illumination, as shown in Fig.\ref{Fig1} d. 

{To evaluate the detection characteristics, we performed direct detection measurements of the graphene detector using a backward-wave oscillator (BWO) as the radiation source. Importantly, we observe zero-bias photovoltage, as expected. The device was illuminated at a frequency of 129.2 GHz with an incident power of 4 mW at the source. Considering optical path losses, the effective power incident on the detector was estimated to be approximately 2 mW. The voltage responsivity of the graphene detector, defined as $r_V = V_{ph}/P_{i}$, where $V_{ph}$ is the measured photovoltage and $P_{i}$ is the incident power, was found to be about 5 V/W under zero-bias conditions. When a bias current was applied to the graphene channel, the responsivity increased to nearly 20 V/W (see Fig.~\ref{Fig2}, b).}

{Another key characteristic of the detector is the noise-equivalent power (NEP), which quantifies the minimum detectable signal power constrained by the noise level. NEP is defined as $NEP = s_V/rV$, where $s_V$ is the noise spectral density and $r_V$ is detector responsivity. Unlike responsivity, NEP is independent of the measurement mode (photovoltage or photocurrent) and thus serves as a more universal figure of merit.
We measured the noise spectral density of our photodetector using selective nanovoltmeter Unipan 233. In the zero-bias regime, the measured noise spectral density agrees well with the expected thermal noise level of the graphene channel $\sqrt{4kTR}$ (see Supporting Information), while the noise increases as expected with the applied current bias. }

{ The minimum achieved NEP of the detector is 0.56 nW/$Hz^{0.5}$ at a current bias of 5 $\mu$A, while at zero bias regime the NEP is 0.68 nW/$Hz^{0.5}$, as shown in Fig.~\ref{Fig2}, b. Although this value is approximately one order of magnitude higher than the best reported room‑temperature graphene detectors~\cite{Ludwig2024,Soundarapandian2026}, the performance of our device can be improved through gate‑voltage tuning.}

{To further characterize the detector properties, we performed frequency-dependent measurements of antenna-coupled graphene detector in 100-200 GHz frequency range (see Fig.~\ref{Fig2} c). These measurements agree well with the calculated antenna parameters under a 1~k$\Omega$ load (see Fig.~\ref{Fig2} c and d).}

Subsequent high-frequency measurements were performed with zero current applied to the channel.

\section*{High-frequency THz measurements}

All frequency-dependent measurements were performed at room temperature and in the zero-bias regime. To achieve high‑frequency modulation of the THz radiation, we employed a heterodyne scheme using two backward wave oscillators (BWOs) — or alternatively, one BWO and a frequency multiplier driven by an RF generator. The two sources generated slightly different frequencies that were combined through a beamsplitter and focused onto the graphene detector (see Fig.\ref{Fig3} a). 
% We fixed the frequency of $\rm{BWO_1}$ at 126.5 GHz while tuning the second THz source from this frequency up to 153 GHz.

The output signal from the graphene photodetector was amplified by a room-temperature high-frequency amplifier and then fed into an electrical spectrum analyzer (ESA), where both the signal and noise floor were recorded at the intermediate frequency ${\rm IF} = f_{\rm THz_1}-f_{\rm THz_2}$. A Golay cell was used to monitor the variation of the THz source power during frequency tuning. We measured the photosignal as a function of IF up to 43 GHz, limited by our ESA.
% Optionally, a bias‑tee was inserted between the graphene detector and the amplifier to enable simultaneous DC biasing/measuring and AC signal measurement in the graphene channel. 

\section*{Results and Discussion}

The measured signal-to-noise ratio (SNR) as a function of IF is shown in Fig.~\ref{Fig3} d. Despite some variations, the SNR exhibits no decreasing trend, indicating that the true bandwidth of our packaged graphene substantially exceeds 43 GHz. We proceed to discuss the principal engineering solutions that enabled the flat sub-THz response up to 43 GHz modulation frequency.

The first innovation lies in the use of a high-impedance antenna instead of attempting to reduce the graphene dc resistance to 50 Ohm for RF line matching. Indeed, integration of sub-THz antennas with high-frequency graphene detectors presents a nontrivial challenge. While antennas are essential for radiation concentration, they typically introduce losses that severely constrain the bandwidth. In Ref.~\cite{Soundarapandian2026} an inverse relationship between device responsivity and bandwidth was established, attributed to losses introduced by the antenna structure necessary to enhance the photoresponse. Our results show that this limitation does not apply to graphene-coupled double-slot antennas. The ability to couple high-resistivity objects to metal slots can be understood by 'expulsion' of terahertz electric field from metal plate into the slot, accompanied by the enhancement in the field magnitude.

The second innovation lies in use of geometrically structured contacts to achieve zero-bias detection. This simple structure avoids the need for complex approaches - such as $p-n$ junctions induced by split gates~\cite{Titova2023,Soundarapandian2026} or dissimilar contact metals~\cite{Cai2014}. A straightforward geometric asymmetry, introduced by a tooth‑like design, suffices to generate a strong zero‑bias photoresponse, resulting in an energy‑efficient and easily fabricated detector. 

The selected detector architecture with geometrically structured contacts can also be beneficial for achieving short response time, in comparison with split gate structures. The origin of effect can be understood by considering the graphene channel as a distributed $RC$-circuit. The propagation of thermoelectric photocurrent from a photoactive junction to either source or drain contact takes finite time proportional to contact-junction separation~\cite{Svintsov_RF_contact_junction}. This time is maximized if the photocurrent is generated between split gates in the middle of the channel. Conversely, this time is minimized if metal-graphene junction in immediate vicinity of the source generates the photocurrent. This result applies not only to graphene-based detectors, but to other detector architectures with 2D channels~\cite{Muravev2012e}. Removal of rectifying gates and transition to contact-based rectification should be beneficial for reduced response time~\cite{Muravev-response_times}. 

% We demonstrate an antenna-coupled, fully packaged sub-THz graphene detector with a response time of less than 40 ps.

% In graphene, the maximum photoresponse typically occurs near the charge‑neutrality point, where the sheet resistance is high, posing challenges for impedance matching with RF lines. Common approaches to address this involve the use of interdigital grid contacts or electrical doping to lower the channel resistance. However, the former requires complex fabrication and constrains device geometry, while the latter often leads to reduced sensitivity. 

% Our results demonstrate that neither the antenna design, the presence of a back-gate, nor the high resistance of the graphene channel limits the bandwidth of sub-THz graphene photodetectors.

Accurately assessing the true performance of such high-speed devices requires careful design of electrical connections and packaging to minimize parasitic delays. We therefore emphasize the importance of fully packaged O/E detector measurements. While optical‑to‑optical characterization methods are suitable for assessing modulators, they are insufficient for evaluating the final implementation of photodetectors. High‑frequency measurements using probe stations can reveal the intrinsic performance of on‑chip metallization; however, they overlook additional losses introduced by external wires, connectors, and packaging. Although bandwidths up to 40 GHz have been demonstrated on probe stations with RF probes~\cite{Soundarapandian2026}, such setups are impractical for real‑world deployment. This limitation underscores the benefits of the packaged detector approach presented here. For practical applications, a straightforward and robust method of extracting the electrical photosignal is essential. Our device is designed accordingly: wired connections to standard RF connectors enable direct integration with external readout electronics, providing a compact, turnkey detector solution with a box‑mounted readout channel.

Importantly, these graphene-based detectors are readily scalable to large-area CVD wafers. Ultrafast carrier dynamics in graphene - observed not only in high-quality exfoliated samples but also in CVD-grown material~\cite{Oum2014,Soundarapandian2026} - underscore their strong potential for commercial THz detection applications.

% Although the intrinsic photoresponse of graphene has been proven to be extremely fast, intrinsic carrier dynamics alone are insufficient to realize a high‑speed detector. In practice, the overall photoresponse involves three sequential processes: (1) photo-carrier generation within the material, (2) conversion of these carriers into a photovoltage or photocurrent at the metal contacts, and (3) extraction and collection of the photovoltage by the external electrical circuitry.

\section*{Conclusions} 
To conclude, we present a fully packaged zero-bias sub-THz graphene detector with a bandwidth exceeding 43 GHz, limited by our measurement equipment. To the best of our knowledge, this represents the highest bandwidth reported to date for antenna-coupled graphene THz detectors. This was enabled through impedance matching of the high-impedance THz antenna to the graphene channel and proper engineering of the high-frequency electrical circuitry surrounding the sample. Zero-bias operation was achieved through the geometric design of graphene metal contacts: one electrode featured a tooth-shaped structure that enabled asymmetric field enhancement in graphene. These graphene photodetectors can be readily adapted across the entire THz range through antenna redesign and hold great promise for enabling high-speed 6G communication systems.

\section*{Methods} 

{\textbf{Sample fabrication.} Graphene and hBN flakes were obtained by mechanical exfoliation from bulk crystals and then assembled into heterostructures using a standard dry-transfer technique. The resulting hBN/Gr/hBN stack was then transferred onto a Si/Si$\rm{O_2}$ substrate with a 300 nm thick oxide layer. Prior to the transfer, a partial antenna structure had been fabricated on the substrate by optical lithography followed by electron-beam evaporation of 3 nm of Cr and 80 nm of Au, leaving the central region free of metal for the placement of the heterostructure. The device geometry was defined by electron-beam lithography (EBL). The heterostructure was patterned via reactive ion etching (RIE), using S$\rm{F_6}$ to etch hBN and $\rm{O_2}$ to etch graphene, leaving a defined graphene channel in the central region. Additional antenna sections (3 nm Cr / 80 nm Au) were then deposited by EBL and electron-beam evaporation to simplify the pattern for subsequent contact formation. Electrical contacts to graphene area were defined in a final EBL step, with the top hBN layer locally removed by RIE (S$\rm{F_6}$) to expose the graphene. Finally, Cr/Au (3 nm/80 nm) contacts were deposited by electron-beam evaporation to complete the device.

}

{\textbf{Photocurrent simulation.}
Electromagnetic simulations were implemented using the Microwave Studio software package, employing a single-frequency finite element solver.\\ 
The photothermoelectric response was modeled using the continuity equation combined with a microscopic expression for the electron current, which includes both diffusion due to temperature gradients and drift caused by electric fields. This results in the following equation for the light-induced electric potential $\varphi(\mathbf{r})$:
\begin{equation}
\label{eq-thermoelectric}
    (\nabla, \alpha_e(\mathbf{r}) \nabla T(\mathbf{r}) 
    - \sigma_e(\mathbf{r}) \nabla \varphi(\mathbf{r})) = 0.
\end{equation}
In this expression, $\alpha_e(\mathbf{r})$ is the thermal diffusivity and $\sigma_e(\mathbf{r})$ is the electrical conductivity.\\ 
Thermal transport was described by the heat conduction equation governing the electron temperature distribution $T(\mathbf{r})$:
\begin{equation}
\label{eq-thermal}
    -(\nabla, \chi_e(\mathbf{r}) \nabla T(\mathbf{r})) 
    = -\frac{C_e}{\tau_\varepsilon}\,(T(\mathbf{r}) - T_0) 
    + \frac{1}{2}\sigma_{\rm opt} |\mathbf{E}|^2.
\end{equation}
Here, $\chi_e$, $C_e$ and $\tau_\varepsilon$  denote the electronic thermal conductivity, heat capacity and the energy relaxation time associated with interactions with substrate phonons, respectively. $T_0$ is the equilibrium temperature. The final term represents Joule heating induced by the electromagnetic field. The quantities $\alpha_e$, $\chi_e$, and $\sigma_e$ are related via the Wiedemann--Franz and Mott relations, leaving only one independent parameter.\\
The calculations in the Fig. 1c are made with the Schottky barrier length $l_{\rm J} = 50 \text{ nm}$ with a stepwise change in conductivity and the characteristic length of thermal relaxation $l_T = \sqrt{\chi_e \tau_\varepsilon / C_{e}} = 100 \text{ nm}$.}

\textbf{Direct photovoltage measurements.} The detector characterization in the direct regime was performed using a BWO source operating over frequencies $f \approx$ 90–260 GHz. The optical path included two THz lenses - one for collimation and another for focusing the beam onto the sample. The linearly polarized radiation was oriented horizontally, perpendicular to the antenna slots. A beam splitter directed part of the beam to a pyroelectric detector to monitor BWO power during frequency sweeps. The graphene photovoltage was recorded simultaneously and normalized to the measured BWO power to obtain the  $V_{ph} (freq)$ dependence.

\textbf{Responsivity and NEP estimations.} Responsivity and NEP were determined using a BWO source calibrated at 
f = 129.2 GHz. The BWO output power was 4 mW, delivered to the collimating lens through WR6 waveguides. In free space, a second lens focused the THz beam onto the sample. The distance between the BWO and the sample was approximately 13 cm. Optical path losses were estimated at -3 dB, giving an incident power of $P_i = 2$ mW. The entire THz beam was assumed to illuminate the Si lens with the sample mounted at its focus; thus, the full beam was considered to irradiate the sample. 

The photovoltage $V_{ph}$ was measured using a sourcemeter Keithley 2450 by recording I–V curves with and without THz illumination, then averaging, and subtracting one from the other. The voltage responsivity was calculated as $r_V = V_{ph}/P_i$.

To determine the noise equivalent power, defined as $\mathrm{NEP} = V_n / r_V$, we measured the voltage noise spectral density $V_n$ (see Supporting Information) at a central frequency of $f_0 = 100\,\mathrm{kHz}$, well above the $1/f$ noise corner, within a bandwidth of $\Delta f = 5\,\mathrm{kHz}$ using a selective nanovoltmeter (Unipan 233). The NEP was then obtained from $NEP = V_n/\sqrt{\Delta f}/r_V$

\textbf{Heterodyne bandwidth measurements.} To measure the response time of our graphene detector, we implemented a heterodyne detection scheme in which two terahertz (THz) beams were mixed on the sample, generating a photosignal at the intermediate (difference) frequency. A backward-wave oscillator (BWO) operating at a fixed frequency served as the local oscillator (LO), while a Keysight N5191A signal generator combined with a ×12 frequency multiplier provided the frequency-tunable radio-frequency (RF) source. The detector output was connected to a Ceyear 4051G electrical spectrum analyzer through a low-noise amplifier. Two amplifier sets were utilized for different intermediate-frequency (IF) ranges: one for IFs below 13 GHz (ZFL-1000LN+ or ZWA-183W) and another covering the 2–50 GHz range (MWAVE LNA-0250-5005).
%  ZFL-1000LN+ or ZWA-183W 

\vspace{1em}

% \section*{Acknowledgements}

\section*{Data availability}
All data supporting this study and its findings are available within the article and its Supplementary Information or from the corresponding authors upon reasonable request.

\section*{Competing interests}
The authors declare no competing interests.

% \section*{Author contributions}
% D.A.B. conceived and supervised the project. L.E., A.L.S., E.T., I.M. and I.G. performed photoresponse measurements. S.J., N.K., M.K., Y.W., and V.D. contributed to the fabrication of  devices. D.A.S. and P.A.P. provided theory support. L.E., A.L.S. analyzed the data with input from D.A.B.. L.E. and D.A.B. wrote the paper with input from all authors. G.G. and K.S.N contributed to discussion on thermoelectricity and THz photoresponse measurements. T.T. and K.W. provided high quality hBN crystals.  

\bibliography{Bibliography.bib}

\newpage
\setcounter{figure}{0}
\renewcommand{\thesection}{}
\renewcommand{\thesubsection}{S\arabic{subsection}}
\renewcommand{\theequation} {S\arabic{equation}}
\renewcommand{\thefigure} {S\arabic{figure}}
\renewcommand{\thetable} {S\arabic{table}}

\end{document}